%% file: main_survey.tex
\documentclass[9pt,twocolumn,twoside]{pnas-new}

\templatetype{pnasresearcharticle} 

\usepackage{multirow}
\usepackage{rotating}
\usepackage{kotex}
\usepackage{pifont}
\usepackage{subfigure} 

\graphicspath{{./FIG/}}

\newcommand{\cmark}{\ding{51}}%
\newcommand{\xmark}{\ding{55}}%
\newcommand{\eat}[1]{}
\newcommand{\eg}{\emph{e.g.}}
\newcommand{\ie}{\emph{i.e.}}

\newenvironment{myitemize}
{ \begin{itemize}[label=$\bullet$]
		\setlength{\itemsep}{0pt}
		\setlength{\parskip}{0pt}
		\setlength{\parsep}{0pt}     }
{ \end{itemize}                  } 

\title{Universal Components of Real-world Diffusion Dynamics based on Point Processes}

\author[a*]{Minkyoung Kim}
\author[a]{Raja Jurdak}
\author[b]{Dean Paini} 

\affil[a]{Data61, Commonwealth Scientific and Industrial Research Organisation (CSIRO), Pullenvale, QLD 4069, Australia}
\affil[b]{Health \& Biosecurity, Commonwealth Scientific and Industrial Research Organisation (CSIRO), Black Mountain, ACT 2601, Australia}

\leadauthor{Kim} 
\correspondingauthor{* Corresponding Author. E-mail address: minkyoung.kim@data61.csiro.au (M. Kim).}


\begin{abstract}
Bursts in human and natural activities are highly clustered in time, suggesting that these activities are influenced by previous events within the social or natural system. 
Bursty behavior in the real world conveys information of underlying diffusion processes, which have been the focus of diverse scientific communities from online social media to criminology and epidemiology. 
However, universal components of real-world diffusion dynamics that cut across disciplines remain unexplored.
Here, we introduce a wide range of diffusion processes across disciplines and propose universal components of diffusion frameworks. We apply these components to diffusion-based studies of human disease spread, through a case study of the vector-borne disease dengue.
The proposed universality of diffusion can motivate transdisciplinary research and provide a fundamental framework for diffusion models.
\end{abstract}

\begin{document}

\verticaladjustment{-2pt}

\maketitle
\thispagestyle{firststyle}
\ifthenelse{\boolean{shortarticle}}{\ifthenelse{\boolean{singlecolumn}}{\abscontentformatted}{\abscontent}}{}

\section{Introduction}
\label{sec:intro}
\input{010intro}

\section{Bursty Behaviors in the Real World}
\label{sec:bursts}
\input{020bursts}	

\section{Universal Effects on Real-world Diffusion Dynamics}
\label{sec:factors}
\input{030factors}

\section{Diffusion Frameworks with Point Processes}
\label{sec:framework}
\input{040framework}

\section{Case Study: Dengue Spread}
\label{sec:casestudy}
\input{050casestudy}

\section{Discussion}
\label{sec:discussion}
\input{060discussion}

\section{Conclusion}
\label{sec:conclusion}
\input{070conclusion}


\bibliographystyle{pnas-new}
\bibliography{ref_survey}

\end{document}

%% file: 010intro.tex

Diffusion processes have received increasing interest from researchers due to their ability to characterize a broad range of real-world situations, from the spread of information in physical~\cite{uzzi1997social,schilling2007interfirm,fleming2007small} and online~\cite{cha2009flash,kim2012eventdiffusion,cheng2014can} networks, to the spread of electronic viruses in computer networks~\cite{newman2002email,balthrop2004technological,wang2009understanding}, to the transfer of stress in earthquakes~\cite{ogata1988statistical,stein1999role,nanjo2007decay}, and the spread of pathogens across a population~\cite{morris1995spread,morse2001factors,read2008dynamic}. 
This real-world diffusion is often presented as a rise and fall of a number of cascading events over time, which is led by behavioral changes of individuals influencing each other.
Examples include the resharing of information exposed to online contacts~\cite{cha2009flash,kwak2010twitter,kim2012eventdiffusion,cheng2014can}, the purchase of a new consumer product recommended by friends~\cite{leskovec2006patterns,leskovec2007dynamics}, the contagion of an infectious disease via direct~\cite{marks2006estimating} or indirect~\cite{bhatt2013global,shahzamal2017social} contacts with infected people or animals, frequent crimes along hotspots~\cite{short2010nonlinear,mohler2011self}, or aftershock sequences in neighboring regions of the seismic center~\cite{ogata1988statistical,ogata1991some,stein1999role}.

A common approach to quantitatively understanding such collective behavior is to characterize their temporal features. 
More specifically, researchers often approximate the inter-activity or inter-event times of human (\eg, communication) or natural (\eg, earthquake aftershock) behavioral sequences by a non-Poisson distribution which captures bursts of intensive activity (bursty behavior) separated by long-term inactivity~\cite{gardner1974sequence,barabasi2005origin,vazquez2006modeling}, not allowed in a uniform distribution.
This implies that the decision processes of individual human activities or the emergent events in cascades are closely related and thus highly clustered~\cite{mohler2011self,hu2012were}.
That is, the timing of these discrete events conveys information of the underlying processes driving diffusion through a network.
Accordingly, uncovering the mechanisms of clustered inter-event times has attracted significant interest among diverse research areas in order to better understand the dynamics of collective bursts, to predict future diffusion trends, and to establish efficient strategies that promote or limit the diffusion process.

As a mathematical tool, the concept of a point process has been widely used as the fundamental framework of a diffusion model, since it realizes any generic bursty behavior with a series of point events on the real line either in time or space~\cite{doob1953stochastic,cox1965theory,daley2007introduction,snyder2012random}. 
The broad applicability of point processes enables the characterization of a wide range of diffusion scenarios in real world networks.
In addition, a point process regards events as continuous arrival processes, which is beyond deterministic approaches based on expensive extraction of exhaustive features as inputs to a simple classifier, or time series classes of functions~\cite{suh2010want,kupavskii2012prediction,bhatt2013global,gardner2013global}. 
Finally, a point process is a model-driven approach estimating non-linear dynamics with parametric assumptions on underlying diffusion processes. 
Alternative model-free approaches that are independent of any assumptions have been advantageous to directly infer the causal relationships between discrete events by using information-theoretic measures~\cite{victor2006approaches,ver2013information,kim2016macro}, yet these approaches rarely provide a rich context on underlying diffusion mechanisms, in contrast to interpretable model-driven approaches.

As a result of these benefits, modeling the diffusion dynamics of individual items has been widely attempted using point process approaches.
Examples include predicting final retweet counts of individual tweets in Twitter~\cite{gao2015modeling,zhao2015seismic}, the repeated uses of textual phrases across social media~\cite{gomez2013modeling}, citation volumes of either a single publication~\cite{wang2013quantifying,shen2014aaai}, a research topic~\cite{kim2017social}, or a field~\cite{sinatra2015century} in science, and the occurrences of earthquakes~\cite{ogata1988statistical,ogata2003modelling}.
Despite the diversity in application domains for diffusion studies, we argue that there are major factors of the underlying processes that are universal and application-agnostic, which have not been sufficiently studied by discipline-focused research communities.

In this paper, we aim to: (1) disclose these common major factors of diffusion processes across different disciplines and propose a taxonomy of universal effects on real-world diffusion dynamics, (2) compare diffusion models based on point processes, in accordance with the disclosed universal effects, and (3) propose a high-level sketch of universal components for a disease diffusion framework as a case study.
We expect that the universal effects of diffusion are interdisciplinary and applicable to a broad range of real-world scenarios, moving towards a more general framework that can incorporate unique scenario-specific contexts.

In the rest of this paper, we first introduce diverse collective bursts in the real world, based on which we then uncover the common factors of underlying diffusion processes across different research areas and propose a taxonomy of the universal effects on real-world diffusion.
We then explain the background of point processes and compare diffusion models with point process approaches in terms of the proposed taxonomy. 
Subsequently, we identify human disease diffusion  based on point processes as an under-explored area and introduce a case study of applying universal diffusion components to a disease diffusion framework. 
This is followed by a discussion of the key design considerations of a diffusion framework, and the future challenges.

%% file: 020bursts.tex

With the help of recent advancements in social sensing~\cite{aggarwal2013social}, diverse human activity records can now be collected at unprecedented scale and resolution.
The large scale and diversity of human lifelogs via social sensors (\eg, activity records via tracing applications on mobile devices such as smart phones) provide a wide spectrum of bursty behavior sequences in time and space enabling the capture and analysis of underlying diffusion processes in the real world~\cite{kim2017social}.
In this section, we introduce various bursty events observed in different research areas in more detail, and we highlight the common factors that have brought the collective bursts in the next section.

\subsection{Bursts of Collective Behavior}
Diffusion patterns are often presented with a rise and fall of the number of relevant events over time. 
In social media, these event bursts appear as recurrent citation curves of web posts~\cite{cheng2014can,cheng2016cascades}, a sudden increase in new connections due to the cascading reshares of a post~\cite{myers2014bursty,farajtabar2015coevolve}, or simultaneous propagation of breaking news across different types of social media~\cite{kim2015dynamics}. 
Such bursty behavior goes beyond online social media and can be observed as abrupt sales growth of consumer durables~\cite{bass2004comments}, rapid increase in the number of protesters against political movements~\cite{kim2013modeling}, or recurrent citation curves of journal articles in science~\cite{wang2013quantifying,shen2014aaai}. 
Disease spread in humans and animals also shares similar patterns, such as the spread of air- or vector-borne viruses in populated areas~\cite{kuno1995review,morse2001factors}, foot and mouth disease among farm livestock~\cite{gerbier2002point,Diggle2006,boender2007local}, and avian influenza in poultry~\cite{Boender2007maps}.

Because these events are not independent of each other, a common feature that cuts across specific applications is that the time intervals between continuous events are not uniformly distributed but clustered in a non-Poisson fashion~\cite{gardner1974sequence,barabasi2005origin,vazquez2006modeling}.
That is, diffusion processes are influenced by previous events, collectively leading to bursts of intensive activity separated by long term inactivity, \ie, clustered inter-event times~\cite{mohler2011self,hu2012were}.

\subsection{Attention Economy}
Individual diffusion items compete for costly attention across a social system, leading to highly skewed diffusion sizes of different scales.
Such competing processes for attracting collective attention are known as the attention economy~\cite{falkinger2007attention}.

In information diffusion contexts in online spaces, the competition for attention has been boosted by record levels of user-generated content~\cite{kim2015dynamics}.
For instance, a growing list of online networking services (\eg, Twitter, Facebook, Youtube) and social media aggregators of news feeds or social networks (\eg, Flipboard, Google Reader, TweetDeck) have enabled users to create and share an unlimited amount of content in different formats, such as microblogs~\cite{kwak2010twitter,cheng2014can,cheng2016cascades}, images~\cite{van2007flickr,lerman2007social,nov2010analysis}, videos~\cite{cha2009flash,crane2008robust,wang2012propagation}, hyperlinks~\cite{myers2012information,kim2013modeling}, social events~\cite{kim2012eventdiffusion,kim2013direct}, and even short text phrases (memes)~\cite{gomez2013modeling}. 
Similarly, bursts of disease outbreaks have become more common with globalisation and the increased regional and inter-country movement of people, goods and animals~\cite{morse2012prediction,jurdak2015autonomous}. 

Accordingly, diverse information items propagate by the sharing behaviors of online users~\cite{cheng2014can} across a single social site~\cite{kwak2010twitter,myers2012information,gao2015modeling,zhao2015seismic}, across multiple sites~\cite{gomez2010inferring}, and across different types of social media (\eg, mainstream news, social networking sites, blogosphere)~\cite{kim2013modeling,kim2014trends}. 
When it comes to scholarly publications, scientific innovation emerges via citing behaviors across a research field~\cite{kim2014socialcom} and across disciplines in science~\cite{sinatra2015century,van2015interdisciplinary}.
In epidemics, disease outbreaks span local, inter-state, and inter-country transmissions, which is affected by nationwide human mobility~\cite{jurdak2015understanding}, virus importation via international travelers~\cite{bhatt2013global,gardner2013global}, and introduction of non-native infectious species via cargo transportation from endemic regions~\cite{kuno1995review,crowl2008spread}. 

Cascade sizes have been characterized with a heavy-tailed distribution~\cite{leskovec2006patterns,bakshy2011everyone,kim2012eventdiffusion}.
The varied scales of cascade sizes imply that individual items experience competing processes for collective attention, leading to popularity disparity in a social media~\cite{wu2007novelty,yang2011patterns,dow2013anatomy,leskovec2009meme,gao2015modeling,zhao2015seismic} or attention inequality in scholarly publications~\cite{wang2013quantifying,shen2014aaai}.
For epidemics, contagious diseases go through spatial expansion processes via movement of infectious humans or animals, leading to risk variation across a population~\cite{gerbier2002point,Diggle2006,boender2007local,bhatt2013global,gardner2013global}.
That is, some diffusion items become widely popular or infectious, while others are quickly forgotten or attenuated. 

\subsection{Exogenous vs Endogenous Bursts}
The lack of prior knowledge about the origin of a burst has motivated researchers to understand the interplay between external shock and internal dynamics (self-organization) in complex systems~\cite{roehner2004response} and how they drive bursts.
The branching structure of exogenous (external) and endogenous (internal) likelihood of the burst origin has been considered in early social science such as ``innovators'' (via external influences), ``early adopters'', ``early majority'', ``late majority'', and ''laggards'' (detailed segmentation of adopting behavior via internal influences)~\cite{rogers1962diffusion} and in marketing such as ``innovation'' and ``imitation'' for each likelihood~\cite{bass1969}.

\begin{figure*}[t]
	\centering
	\includegraphics[width=1\textwidth]{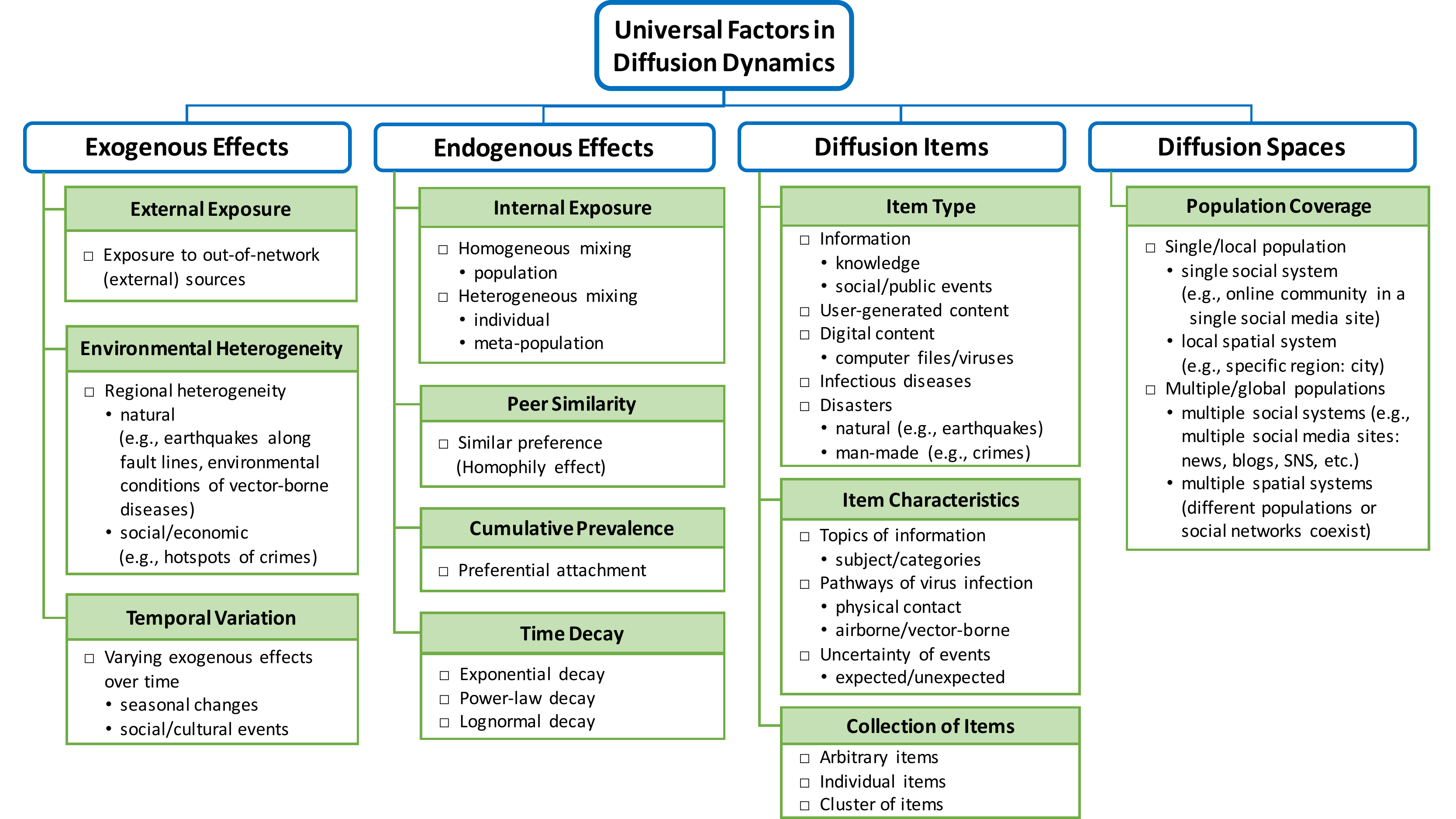}
	\caption{A taxonomy of universal factors in real-world diffusion dynamics}
	\label{fig:taxonomy}
\end{figure*}

More recently, large scale social media data has enabled researchers to investigate the effects of these two categories on diffusion patterns.
Bursty attention initiated by stronger exogenous than endogenous effects drives asymmetric diffusion curves, while endogenous bursts lead to symmetric ones~\cite{crane2008robust}.
For example, an unexpected disastrous event (exogenous) such as a tsunami brings about abrupt public attention (asymmetric diffusion), while an expected film release (endogenous) such as the release of a Harry Potter movie is paid precursory mass appeal (symmetric diffusion).
In terms of news category, synchronous (simultaneous) or asynchronous (nonparallel) diffusion patterns have been discovered across different types of social media~\cite{kim2015dynamics}.
For instance, controversial political topics such as political protests in the Middle East and multiculturalism failure exhibit sharply growing diffusion patterns concurrently across mainstream news sites, blogs, and social networking sites, while entertainment topics such as celebrity news show different diffusion patterns across the media. 
Similar dynamics are evident for disease spread. 
Seasonal infections, such as flu-like illnesses, tend to exhibit symmetric diffusion curves with gradual buildup of infections in colder months and a gradual decrease as the weather gets warmers~\cite{zhang2017flu}. 
Imported epidemics, such as the outbreak of Zika in South America, exhibit a sharp rise in infections followed by gradual reductions as the medical community responds with vaccines and prevention measures~\cite{paho2017}.

All in all, bursty behaviors convey information of diffusion dynamics, triggered by exogenous and/or endogenous influences over a network, exhibiting distinct diffusion patterns.
In the next section, we discuss common major effects on real-world diffusion in more detail.

%% file: 030factors.tex


Real-world diffusion processes have been mathematically modeled in order to understand underlying mechanisms governing emergent collective behavior. 
This has been accomplished by incorporating different diffusion factors as major parameters into  diffusion models.
Before discussing the frameworks of prior diffusion models in the next section, we first examine common major factors of collective bursty behaviors across different research areas such as computer science, social science, statistical physics, seismology, criminology, marketing, finance, and epidemics.
Based on the common factors across disciplines, we propose a taxonomy of universal effects on real-world diffusion dynamics, as shown in Figure~\ref{fig:taxonomy}.

We categorize a broad range of diffusion factors into four major elements: (1) exogenous effects that characterise influences from outside the (social) network under consideration; (2) endogenous effects that describe the internal dynamics of the target (social) system; (3) diffusion items that specify what is spreading and what are its features; and (4) the diffusion space that defines the boundary of diffusion dynamics.

\subsection{Exogenous Effects}
Out-of-network effects on diffusion are called {\it exogenous effects} or {\it external influences}, and can be characterized by {\it external exposure}, {\it environmental heterogeneity},  and/or {\it temporal variation}.
In the point process approaches~\cite{crane2008robust,simma2010modeling,iwata2013discovering,farajtabar2015coevolve}, this exogenous effect is often called ``background intensity'' or ``baseline intensity'' representing the rate of events occurring without considering the influence of preceding events.

\subsubsection{External Exposure}
Collective behavioral changes can be triggered by external sources outside of a (social) network.
More specifically, accessibility or exposure rate to out-of-network sources can influence diffusion patterns, such as the purchase of a consumer product influenced by advertisements on traditional mass media (\eg, TV, newspaper) in marketing~\cite{kumar2002multinational,bass2004comments}, the adoption of new information via news feeds or other external sources (\eg, mainstream news sites, blogs, social networking sites) in online social media~\cite{myers2012information,kim2013modeling,gomez2013modeling}, and the introduction of an exotic virus by international travelers from endemic regions~\cite{morse2001factors,wilder2008geographic,bhatt2013global}. 

Often, external influences have either been ignored~\cite{romero2011hashtag,wang2013quantifying,shen2014aaai}, or simply approximated as a constant value for all different diffusion items~\cite{iwata2013discovering}.
However, in online social media, a large proportion (30\%) of Twitter users are exposed to a trending topic, influenced by external information sources such as mainstream news, blogs, or personal homepages rather than by internal contacts (followees)~\cite{myers2012information}.
For disease spread, new outbreaks can be initiated by travellers importing the disease from regions in which the disease is endemic into other regions where the disease is not normally present~\cite{kuno1995review,morse2001factors,bhatt2013global,gardner2013global}. 
As human mobility increases, the fluxes of travellers from disease-endemic into non-endemic regions are likely to increase.
Analysis that focuses only on the latter regions without considering imported cases is therefore incomplete. 

Such a large or increasing proportion of external exposures are clearly not negligible, as often diffusion sources are not limited to local contacts~\cite{myers2012information} and there is a likelihood of direct connectivity between distant social networks~\cite{kim2013modeling}, such as Facebook and Twitter in an online media context, or two geographically neighboring countries in a disease spread context.

\subsubsection{Environmental Heterogeneity}
There are intrinsic environmental effects on diffusion, which may show little change over time, but are spatially heterogeneous.
For example, we more often observe earthquakes along fault lines~\cite{albala2000complex,ogata2003modelling}, disease outbreaks in specific regions satisfying environmental conditions (\eg, temperature, humidity, precipitation) of virus transmission~\cite{bhatt2013global,shahzamal2017social}, or crimes along hotspots~\cite{short2010nonlinear,mohler2011self}.
Such environmental heterogeneity has been incorporated to diffusion models for understanding clustered events in both space and time~\cite{ogata2003modelling,mohler2011self}, but this spatial heterogeneity can be applicable to meta-spaces at an abstract level, such as different social networks in online social media~\cite{kim2015dynamics} and hierarchical regional subpopulations in a country~\cite{colizza2008epidemic} as meta-populations.

\subsubsection{Temporal Variation}
Exogenous effects can change over time, and the range of fluctuations is situational.
For instance, vector-borne diseases such as dengue~\cite{world2009dengue} are affected by seasonal changes, since breeding conditions (\eg, precipitation, temperature) of mosquito vectors vary by seasons, and external exposures to international travelers from the virus endemic regions likely increase during the peak vacation season.
Meanwhile, social and cultural changes also bring time-variant exogenous effects.
For example, breakthroughs in web technologies make diverse information sources highly accessible, leading to increasing external influences on online social networks, while time-evolving economic situations or city planning bring about changes of crime hotspots. 
The background intensity therefore becomes a function of time.

\subsection{Endogenous Effects}

Contrary to the out-of-network effects, the {\it endogenous effects} describe the internal dynamics in a network starting from a group of initial spreaders, such as innovators or early infected individuals~\cite{bass1969,rogers1962diffusion}, to the broader network.
Accordingly, the endogenous effect is also called ``internal influence'', in contrast to external influence, or ``endogenous intensity'', as opposed to background intensity.
This endogenous effect can be characterized by three major sub-effects such as (1) {\it internal exposure}, (2) {\it cumulative prevalence}, and (3) {\it time decay}. 
 
\subsubsection{Internal Exposure}
We often observe {\it topological heterogeneity} of individuals in a social network, such as heavy-tailed or power-law degree distributions~\cite{newman2005power,clauset2009power,newman2010networks}, which leads to unequal peer influences on diffusion.
For instance, previous adopters or already infected individuals with higher degrees (the number of connections in the network) contribute more to spreading information or pathogens to their unadopted or susceptible neighbors, compared to adopted or infected nodes with lower degrees.
That is, the effects of network structure on diffusion~\cite{albert2002statistical,boccaletti2006complex,newman2010networks}, such as cluster density and reachability~\cite {kuperman2001small,schilling2007interfirm}, and degree distributions~\cite{luu2012modeling,kim2013modeling}, are significant.

However, traditional diffusion models~\cite{bass1969,bailey1975mathematical} assume {\it homogeneous mixing} of a target population, where each individual meets every other individual completely at random, \ie, {\it fully mixed}~\cite{newman2010networks}.
These models provide macroscopic views of the overall adopting/infected population patterns by using differential equations, but they disregard the effects of structural network properties on diffusion.
In contrast to such unrealistic assumptions, there have been extensive studies on diffusion models by considering {\it heterogeneous mixing}~\cite{barrat2008dynamical} among individuals~\cite{luu2012modeling,iwata2013discovering,farajtabar2015coevolve,zhao2015seismic}, meta-populations~\cite{colizza2008epidemic,liu2013contagion}, or both~\cite{kim2013modeling,kim2014socialcom}.

Meanwhile, there have been attempts to separate the {\it peer similarity} (\ie, homophily) effect from peer influence in order to obtain more accurate and detailed interpretations of the network effect~\cite{anagnostopoulos2008influence,aral2009distinguishing}.
That is, individuals' behavioral changes (\eg, information sharing, product purchase) are not only due to exposures to previous adopters in their contacts, but also attributed to similar bonds or common personal properties, such as preference, age, gender, and religion~\cite{yavacs2014impact}.

\subsubsection{Cumulative Prevalence}
The quantitative status of an individual diffusion item up to the current time, \ie, {\it cumulative prevalence}, affects the item's future prevalence or popularity, which captures the ``rich-get-richer'' mechanism~\cite{newman2001clustering,barabasi2002evolution}.
That is, diffusion items with higher prevalence are more likely to spread to unadopted individuals in a (social) system, representing the preferential attachment mechanism~\cite{barabasi1999emergence}. 
For instance, a scholarly article is likely to receive new citations proportionately to its total citation volume~\cite{wang2013quantifying,shen2014aaai}. 
When it comes to social media, the subsequent resharing of a single microblog is also likely influenced by its cumulative reshared frequency~\cite{gao2015modeling,zhao2015seismic}.
Similarly, in disease spread, outbreaks of pathogens that have so far been highly infective are more likely to grow faster~\cite{bailey1975mathematical}. 

\subsubsection{Time Decay}
During the entire diffusion process, an individual item is aging over time.
For instance, as time elapses, a microblog loses its popularity in a social media platform~\cite{gao2015modeling,zhao2015seismic}, a scholarly publication experiences its fading novelty in science~\cite{wang2013quantifying,shen2014aaai}, and a pathogen loses its infectiousness over time in epidemics~\cite{zelner2013linking}.

Due to the different aging processes of individual items, the time decay effects are formulated with different time relaxation functions such as exponential decay~\cite{wu2007novelty,mohler2011self,embrechts2011multivariate,iwata2013discovering,parolo2015attention,shahzamal2017social}, power-law decay~\cite{ogata2003modelling,crane2008robust,gao2015modeling,zhao2015seismic,parolo2015attention}, lognormal decay~\cite{wang2013quantifying,shen2014aaai}, or gamma decay~\cite{simma2010modeling} across different application domains.
Additionally, even within the same scholarly publication domain, novelty decay of journal articles has been approximated with different functions between disciplines. 
For instance, for the majority of publications in the STEM fields, the life-cycle of citations are better explained with exponential rather than power-law decays~\cite{parolo2015attention}.  
On the other hand, in the physics area, a single paper's citation decay has been defined with a lognormal function~\cite{wang2013quantifying,shen2014aaai}.
These all imply that time decay patterns are not only dependent on application domains but also influenced by internal dynamics of different subgroups in the same social system.

\subsection{Diffusion Items}
Some prior work on diffusion has only focused on finding influential nodes in the (social) networks~\cite{kempe2003maximizing,ghosh2010predicting}, by quantifying the structural positions of the nodes in order to incorporate topological effects on diffusion.
Their fundamental assumption is that a node's influence is determined by its topological position only, and the strength of influence is constant for arbitrary diffusion items.
However, the nature of diffusion items cannot be ignored, since it does not only provide the context of diffusion but also affect their spreading pathways over the social networks~\cite{romero2011hashtag,kim2015dynamics}.
In this respect, we examine the nature of diffusion items in terms of the type, characteristics, and collection of the items.

\subsubsection{Item Type}
Diffusion items can be categorized into different types such as information, user-generated content, digital content, pathogens, and natural or man-made disasters, as shown in Figure~\ref{fig:taxonomy}.
Different types of items can also be categorized into hierarchical sub-types.
For example, information consists of knowledge (\eg, scientific publications~\cite{wang2013quantifying,shen2014aaai,kim2014socialcom,parolo2015attention}), social events~\cite{kim2012eventdiffusion,kim2013entropy,kim2016macro}, and consumer durables~\cite{kumar2002multinational,bass2004comments} while disasters include natural and man-made disasters. 
Diseases can be infectious and/or transmissible, and even infectious disease can further classed into more granular categories, such as air- and vector-borne diseases. 
Note that the types presented in Figure~\ref{fig:taxonomy} are intended to be indicative of the breadth of diffusion process rather than exhaustive of all possible types of diffusion items.

\subsubsection{Item Characteristics}
Diffusion patterns vary with the specific features of the diffusion item, such as the topics of individual information items~\cite{romero2011hashtag}.
For instance, political topics are largely triggered by  external influence, while entertainment topics are driven more by internal influence such as interpersonal buzz~\cite{myers2012information,kim2013direct}.
More specifically, diverse news topics can be categorized into general news categories such as politics, economy, education, culture, disasters, celebrity, sports, and technology~\cite{romero2011hashtag,kim2012eventdiffusion}.
Public events can be grouped into expected (\eg, film releases) or unexpected events (\eg, natural disasters), driving precursory growth and symmetric decay of individual responses to the events, or abrupt growth and asymmetric decay of responses, respectively~\cite{crane2008robust}.
That is, the topic of information itself has different levels of intrinsic attractiveness forming inconstant information pathways over social networks.

In addition, the way to transfer items also affects the underlying processes. 
For instance, infectious diseases are transmitted from infected to susceptible individuals via physical contacts such as HIV/AIDS~\cite{marks2006estimating} or indirect contacts such as airborne (\eg, influenza)~\cite{brankston2007transmission,shahzamal2017social} and vector-borne diseases (\eg, dengue, chikungunya)~\cite{bhatt2013global,roth2014concurrent}.

\subsubsection{Collection of Items}
As discussed earlier in this section, some prior work attempted to quantify topological effects of individual nodes in the social networks on propagating arbitrary diffusion items~\cite{kempe2003maximizing,ghosh2010predicting}.
However, diffusion items exhibit varied scales of cascade sizes, approximated by a power-law distribution~\cite{kim2012eventdiffusion,kim2014socialcom}, leading to popularity disparity between individual items~\cite{falkinger2007attention} as discussed in the attention economy section.

Accordingly, an individual item has been the focus for modelling its popularity dynamics, such as a hyperlink~\cite{myers2012information}, a single tweet~\cite{gao2015modeling,zhao2015seismic}, memes (short textual phrases)~\cite{gomez2013modeling}, and an individual article in science~\cite{wang2013quantifying,shen2014aaai}.
In addition, a collection of items have been also considered to predict the popularity of relevant items as a super set~\cite{romero2011hashtag,kim2013modeling,kim2014socialcom,kim2014trends}.
For instance, multiple hyperlinks are assigned a single topic by recognizing named entities (\eg, person, organization, place) in the main content of the reference pages, resolving entities with data matching techniques~\cite{chr12}, and classifying documents into relevant topics~\cite{kim2012eventdiffusion}.
Also, a collection of topics are grouped into upper categories in order to obtain higher-level insight on diffusion trends~\cite{romero2011hashtag,myers2012information,kim2013modeling,kim2014trends}.

\subsection{Diffusion Space}
Diffusion space is a fundamental constraint of diffusion processes, since it determines the heterogeneity of populations in terms of diversity of subgroups or social systems, and inter- and intra-interactions between and within subgroups or systems.
In other words, diffusion space defines the boundary of the diffusion process.

\subsubsection{Population Coverage}
As external exposures are far-reaching across distant (social) systems, the boundary of diffusion space is not clear and needs to be constrained for more accurate models of underlying processes.
As discussed in the section on exogenous effect, a large proportion of Twitter users are exposed to external sources~\cite{myers2012information}, which suggests the extension of a diffusion boundary beyond a single social platform in order to obtain extended diffusion dynamics across multiple populations in social media such as blogosphere~\cite{adar2005tracking,leskovec2007pattern},
blogosphere-to-news~\cite{leskovec2009meme,gomez2013modeling}, blogosphere-to-YouTube~\cite{cha2009flash}, or blogosphere-news-SNS~\cite{kim2013entropy,kim2016macro}.
Similarly, knowledge transfers over research communities within a single discipline~\cite{shi2009information,zhu2013bibliometric,kim2014socialcom}, which collectively leads to interdisciplinary scientific innovations~\cite{boyack2005mapping,porter2009science,kiss2010can,parolo2015attention,sinatra2015century}.

Multidimensional event clustering in both space and time demands the consideration of diffusion space including meta-populations whose interactions are covered by human mobility~\cite{jurdak2015understanding} or information sharing~\cite{leskovec2009meme,gomez2010inferring,cheng2016cascades} in the geographical~\cite{colizza2008epidemic,liu2013contagion} or online space~\cite{gomez2013modeling,kim2015dynamics}, respectively.
That is, a meta-population scheme~\cite{barrat2008dynamical} covers single/local population or multiple/global populations according to the boundary of diffusion space.

%% file: 040framework.tex

\begin{sidewaystable*}[tbhp]
	\centering
	\caption{Diffusion models based on point processes are compared from the aspects of the proposed major universal effects on real-word dynamics in Figure~\ref{fig:taxonomy}. }	
	\renewcommand\arraystretch{1}
	\begin{tabular}{@{}>{\centering\arraybackslash}m{.4in} >{\centering\arraybackslash}m{.65in} >{\centering\arraybackslash}m{.65in} >{\centering\arraybackslash}m{.0in} >{\centering\arraybackslash}m{.65in} >{\centering\arraybackslash}m{.6in} >{\centering\arraybackslash}m{.5in} >{\centering\arraybackslash}m{.0in} >{\centering\arraybackslash}m{.85in} >{\centering\arraybackslash}m{.9in} >{\centering\arraybackslash}m{.65in} >{\centering\arraybackslash}m{.73in} >{\centering\arraybackslash}m{.63in}@{}} 
		\toprule
		\multirow{3}{*}{Model} & \multicolumn{2}{c}{Exogenous Effect} && \multicolumn{3}{c}{Endogenous Effect} && \multicolumn{3}{c}{Diffusion Item} & \multirow{2}{*}{Diffusion} & \multirow{2}{*}{Application} \\
		\cmidrule{2-3} \cmidrule{5-7} \cmidrule{9-11}
		& External Exposure &  Environmental Heterogeneity && Internal Exposure & Time \hspace{1in} Decay & Cumulative Prevalence && Type & Characteristics & Collection & Space & Domain \\ 
		\midrule		
		Kum02~\cite{kumar2002multinational} & coefficient of innovation & \xmark && homogeneous mixing & \xmark & \cmark && information (consumer product)  & different types of consumer durables & individual product & multiple (multinational) & marketing\\		
		\midrule 
		Oga03~\cite{ogata2003modelling} & \xmark & background seismicity rate && heterogeneous mixing & power-law & \xmark && natural disaster (earthquake) & stress from the epicenter & arbitrary stress & single \qquad (Japan) & seismology\\		
		\midrule
		Ege10~\cite{egesdal2010statistical} & background rate & \xmark && homogeneous mixing & exponential & \xmark && man-made disaster (crime) & gang rivalry & arbitrary \quad gang rivalry & single \qquad \quad (local region) & \multirow{3}{\linewidth}{\centering criminology}\\
		\cmidrule{1-12}	 	
		Moh11~\cite{mohler2011self} & \xmark & background burglary rate && heterogeneous mixing & exponential & \xmark && man-made disaster (crime) & residential burglary & arbitrary burglary & single \qquad \quad (local region) & \\
		\midrule 	
		Emb11~\cite{embrechts2011multivariate} & immigration intensity & \xmark && homogeneous mixing & exponential & \xmark && stock & extreme prices (positive \& negative) & arbitrary \: price & N/A & finance \\
		\midrule
		Dig06~\cite{Diggle2006} & baseline intensity & \xmark && heterogeneous mixing & \xmark & \xmark && infectious disease (foot \& mouth dz.) & airborne transmission & arbitrary \quad virus & single \qquad \qquad (UK) &  \multirow{4}{\linewidth}{\centering animal infections} \\
		\cmidrule{1-12}		
		Boe07~\cite{Boender2007maps} & \xmark & \xmark && heterogeneous mixing & gamma & \xmark && infectious disease (avian influenza) & contact based transmission & arbitrary \quad virus & single (Netherlands) &  \\	
		\midrule 		
		Cra08~\cite{crane2008robust} & exogenous source & \xmark && homogeneous mixing & power-law & \xmark && user-generated content (video)  & different classes of videos & individual \: video & single \qquad (YouTube) & \multirow{8}{\linewidth}{\centering social services sharing information} \\
		\cmidrule{1-12}
		Sim10~\cite{simma2010modeling} & baseline intensity & \xmark && homogeneous mixing & exponential/ gamma & \xmark && user-generated content (post) & different types of response & individual \: post & single (Twitter/Wikipedia) &  \\
		\cmidrule{1-12}		
		Luu12~\cite{luu2012modeling} & external influence & \xmark && heterogeneous mixing & exponential & \cmark && information \quad (book) & most popular books & individual \: book info. & single (GoodReads) & \\
		\cmidrule{1-12}			
		Iwa13~\cite{iwata2013discovering} & background intensity & \xmark && heterogeneous mixing & exponential & \xmark && information (web page bookmark) & different types of bookmarks (9 tags) & individual bookmark & single (Delicious) & \\
		\midrule 
		Mye12~\cite{myers2012information} & event profile & \xmark && homogeneous mixing & \xmark & \cmark && information (hyperlink) & different categories & individual hyperlink & single \quad (Twitter) & \multirow{8}{\linewidth}{\centering social networking service} \\
		\cmidrule{1-12} 
		Gao15~\cite{gao2015modeling} & \xmark & \xmark && heterogeneous mixing & power-law & \cmark && user-generated content (tweet) & arbitrary tweet & individual tweet & single \quad (Twitter) &   \\
		\cmidrule{1-12}
		Zha15~\cite{zhao2015seismic} & \xmark & \xmark && heterogeneous mixing & power-law & \cmark && user-generated content (tweet) & arbitrary tweet & individual tweet & single \quad (Twitter) & \\ 
		\cmidrule{1-12}
		Far16~\cite{farajtabar2015coevolve} & baseline intensity & \xmark && heterogeneous mixing & exponential & \xmark && user-generated content (tweet) & arbitrary tweet & individual tweet & single \quad (Twitter) & \\
		\midrule		
		Gom13~\cite{gomez2013modeling} & baseline function & \xmark && heterogeneous mixing & arbitrary time shaping func. & \cmark && meme (textual phrase) & different topics & individual meme & multiple \quad (across sites) & \multirow{3}{\linewidth}{\centering online social media}\\
		\cmidrule{1-12}
		Kim13~\cite{kim2013modeling} & external influence & \xmark  && heterogeneous mixing & \xmark & \cmark && social event & news category & individual news topic & multiple (across social media) & \\ 
		\midrule
		Wan13~\cite{wang2013quantifying} & \xmark & \xmark && homogeneous mixing & lognormal & \cmark && knowledge \quad (paper) & field \qquad \qquad (physics papers) & individual paper & multiple \quad (across science) & \multirow{4}{\linewidth}{\centering scholarly publications in science}\\ 		
		\cmidrule{1-12}
		She14~\cite{shen2014aaai} & \xmark & \xmark && homogeneous mixing & lognormal & \cmark && knowledge \quad (paper) & field \qquad \qquad (physics papers) & individual paper & multiple \quad (across science) & \\ 
		\bottomrule 
	\end{tabular}
	\label{tab:models}
\end{sidewaystable*}

As discussed in the introduction, collective bursty behaviors in the real world can be characterized by point processes.
In this section, we begin with a brief introduction to point processes.
We then investigate diffusion models from diverse research areas, whose fundamental frameworks are based on point processes.
Finally, we compare the models from the aspects of the disclosed universal effects on diffusion dynamics in the previous section.

\subsection{Point Processes}
A point process is an ordered set of random variables in time, geographical space, or more general spaces~\cite{doob1953stochastic,cox1965theory,daley2007introduction,snyder2012random}.
We focus on a temporal point process whose realization consists of a set of isolated points on a timeline~\cite{brillinger2002point} (\ie, time-evolving random variables~\cite{kim2015dynamics}) as a wide range of real-world diffusion phenomena can be represented with temporal point processes.

More specifically, a temporal point process is a counting process $\{N(t), t \ge 0\}$, where $N(t)$ represents the number of events that occur up to time $t$, and it can be characterized by a conditional intensity function, $\lambda(t)$ as
\begin{align}
\label{eq:intensity_r}
\lambda(t) 
&= \lim_{\Delta t \rightarrow 0} \frac{P\{N(t+\Delta)-N(t)=1 \mid H_t\}}{\Delta t} \enspace,
\end{align}
where $H_t$ is the history of event occurrences before time $t$.  
That is, the intensity function $\lambda(t)$ can be considered the expected infinitesimal rate of an event in the immediate future, given the observations prior to time $t$~\cite{daley2007introduction}.

\subsubsection{Poisson Processes}
A Poisson process assumes that an event occurs independent of previous events.
When the intensity function $\lambda(t)$ is constant over time, \ie, constant rate of events per unit of time, it is called a homogeneous Poisson process.
Otherwise, it is called a nonhomogeneous Poisson process~\cite{rausand2004system} which allows the changes of the intensity rate over time.

\subsubsection{Hawkes Processes}
In realistic situations, a collection of events are often observed as clusters in time~\cite{barabasi2005origin}, which implies that the intensity is likely dependent on the history of event occurrences. 
A Hawkes process is a non-Markovian extension of the Poisson process or a Poisson cluster process~\cite{moller2005perfect}, which enables clustering the arrival of events~\cite{hawkes1971spectra}.
That is, the intensity is modeled such that an event excites a Poisson process consisting of other events in a series of point events.
Thus, a Hawkes process is also called a self-exciting point process~\cite{hawkes1971spectra,hawkes1974cluster} and is defined as
\begin{align}
\label{eq:hawkes}
\lambda(t) 
&= \mu + \sum_{\{k:t_k < t\}} g(t-t_k) \enspace,
\end{align}
where $\mu$ denotes a baseline intensity, and $g$ is a triggering kernel whose form has been proposed across diverse research areas by incorporating major factors of clustered interevent times.
As shown in Eq.(\ref{eq:hawkes}), event sequences can be divided into background events via a Poisson process and triggered events via self-exciting processes of previous events (\ie, a Poisson cluster process), forming a branching structure of immigrants and descendants~\cite{embrechts2011multivariate}, innovators and imitators~\cite{bass1969,rogers1962diffusion}, or external and internal influences~\cite{myers2012information,luu2012modeling,kim2013entropy}.
As discussed in the previous section of temporal variation, exogenous effects can change over time.
That is, the background intensity, $\mu$ in Eq.(\ref{eq:hawkes}) becomes a time function, $\mu(t)$.

\subsection{Comparisons of Diffusion Frameworks}
One of the primary questions in emergent collective behavior is to predict the future evolution of bursty dynamics.
This has been accomplished by building mathematical models, estimating model parameters, and understanding the underlying mechanisms based on the inferred parameter values.
In this regard, it has been of common interest to predict an individual item's propensity to spread in diverse application domains such as marketing~\cite{bass1969,jain1991investigating,kumar2002multinational,bass2004comments}, seismology~\cite{ogata1988statistical,ogata2003modelling,nanjo2007decay}, criminology~\cite{short2010nonlinear,mohler2011self}, finance~\cite{embrechts2011multivariate,bacry2015hawkes}, social networking services~\cite{myers2012information,gao2015modeling,zhao2015seismic,farajtabar2015coevolve} or social sharing services~\cite{crane2008robust,simma2010modeling,luu2012modeling,iwata2013discovering}, news publications in online social media~\cite{kim2013entropy,gomez2013modeling}, and scientific publications in science~\cite{wang2013quantifying,shen2014aaai,kim2014socialcom}.
 
In the marketing literature, predicting popularity growth of consumer durables has been a key to the successful establishment of marketing strategies~\cite{takada1991cross}.
This was initially attempted by defining the hazard function as a simple linear form of the proportion of the cumulative adoptions, represented with an ordinary differential equation as its fundamental diffusion framework~\cite{bass1969}, based upon the assumption of a fully connected population~\cite{kim2015dynamics}.

In contrast to this unrealistic assumption of homogeneous mixing like traditional epidemic models~\cite{bailey1975mathematical,newman2010networks}, there have been extensions with respect to (1) ``heterogeneity'' of populations such as multinational diffusion of a consumer product from technology-leading to technology-lag or neighboring countries~\cite{kumar2002multinational}, (2) ``structural connectivity'' such as degree distributions~\cite{luu2012modeling,myers2012information} and network density and diameter~\cite{kuperman2001small,schilling2007interfirm}, or (3) both such as information diffusion over heterogeneous social networks as a whole network, by incorporating the diversity of populations and their structural connectivity~\cite{kim2013entropy}, where the hazard functions are represented as probabilistic generative models~\cite{luu2012modeling,myers2012information,kim2013entropy}.

The hazard function captures the effect of the time interval between two discrete events.
It describes a probability distribution of the elapsed time $T$ for an event to occur within the next infinitesimally small time window $\Delta t$ as
\begin{align}
\label{eq:hazard}
h(t) 
&= \lim_{\Delta t \rightarrow 0} \frac{P\{t \le T \le t+\Delta t \mid T \ge t\}}{\Delta t} \notag \\
&= \frac{f(t)}{1-F(t)} = \frac{f(t)}{S(t)} \enspace ,
\end{align}
where $f(t)$ and $F(t)$ are the probability density and cumulative distribution functions respectively, and $S(t)$ $\left( = 1-F(t) \right )$ denotes the survival function.

The hazard function specifies the instantaneous probability of an event (the probability of experiencing the event of interest in the small time interval $[t, t+ dt)$), given no event has occurred before time $t$.
The probability then becomes $h(t)dt$.
As Eq.(\ref{eq:hazard}) shows, the hazard function is finite and nonnegative, and more importantly it determines the probability density function of event timing.
In this regard, prior work on diffusion processes has aimed at inferring the hazard function by incorporating major factors~\cite{bass1969,jain1991investigating,kumar2002multinational,bass2004comments,luu2012modeling,kim2013entropy,gomez2013modeling}, represented with observed and unobserved covariates as explanatory variables, which helps to obtain the probability distributions of interevent times.
Note that the hazard function and the intensity functions are interchangeably used~\cite{prentice1981regression} as the two functions determine the conditional probability distributions of inter-activity times.
Accordingly, the previous models~\cite{bass1969,jain1991investigating,kumar2002multinational,luu2012modeling,myers2012information,kim2013entropy} can be considered nonhomogeneous Poisson processes, which define the hazard rate with covariates embedding major factors of interevent intervals.
When the hazard function is a constant, it corresponds to a homogeneous Poisson process~\cite{kingman1993poisson}.

As nonhomogeneous Poisson processes, generative prediction frameworks have been proposed for predicting animal infection risks between farms~\cite{Diggle2006,Boender2007maps}, individual tweets' retweet counts~\cite{gao2015modeling,zhao2015seismic} and citations of individual articles in the physics~\cite{wang2013quantifying,shen2014aaai,sinatra2015century} and computer science~\cite{shi2009information,kim2014socialcom} areas.
The proposed intensity functions consist of cumulative prevalence (\eg, retweet counts and citation volumes up to current time) and time decay factors.
Regardless of such time-varying intensity of nonhomogeneous Poisson processes, the self-exciting nature of point events is ignored due to the property of statistical independence of interevent times, and thus an event is not allowed to increase the rate of successor events for some period of time~\cite{hawkes1971spectra}.

In this regard, Hawkes processes have also been adopted as the fundamental diffusion frameworks of prediction models in diverse application domains such as predicting co-evolving trends of individual tweets' popularity and changing network structures~\cite{farajtabar2015coevolve}, different types of responses to Twitter messages and Wikipedia revisions~\cite{simma2010modeling}, the bookmarking frequencies of individual web pages~\cite{iwata2013discovering}, the repeated uses of textual phrases (memes) across social media~\cite{gomez2013modeling}, extreme price moves (negative and positive excesses) from stock market index data~\cite{embrechts2011multivariate}, crime occurrences along hotspots~\cite{mohler2011self}, and aftershock sequences in seismology~\cite{ogata2003modelling}.

\begin{figure*}
	\begin{tabular}{c} 
		\hspace{-1em} \addtolength{\subfigcapskip}{8pt} 
		\subfigure[An arrival process of outbreaks]{
			\includegraphics[width=0.48\textwidth]{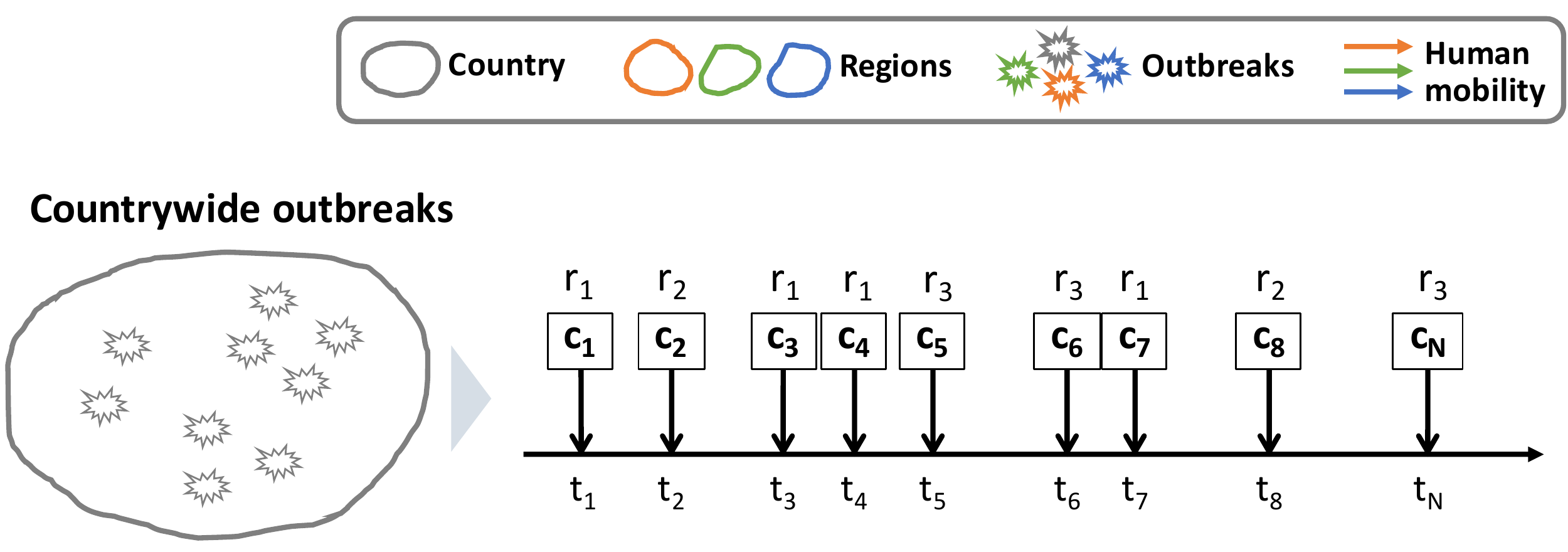}} \vspace{1em}\\ 
		\hspace{-1em} \addtolength{\subfigcapskip}{8pt} 
		\subfigure[Spatial and temporal point processes ]{
			\includegraphics[width=0.48\textwidth]{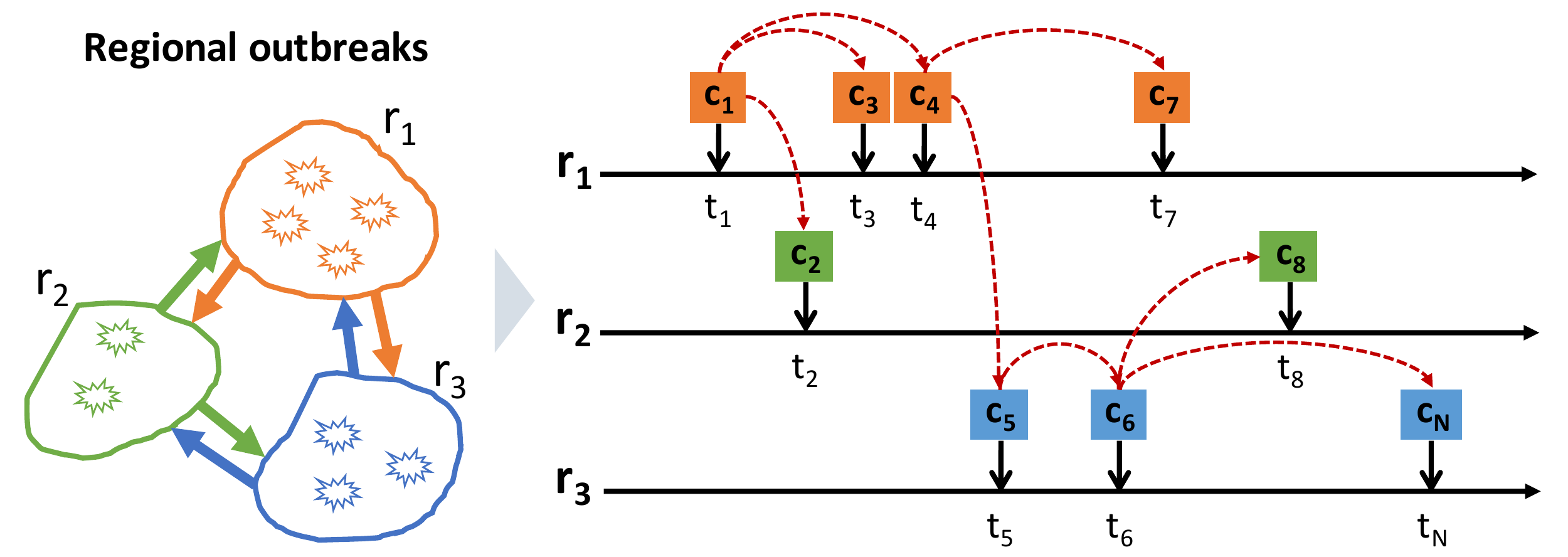}} 
	\end{tabular} \hspace{.1em}
	\begin{tabular}{c} \addtolength{\subfigcapskip}{8pt}
		\subfigure[Conceptual diagram of a diffusion framework]{\includegraphics[width=0.47\textwidth]{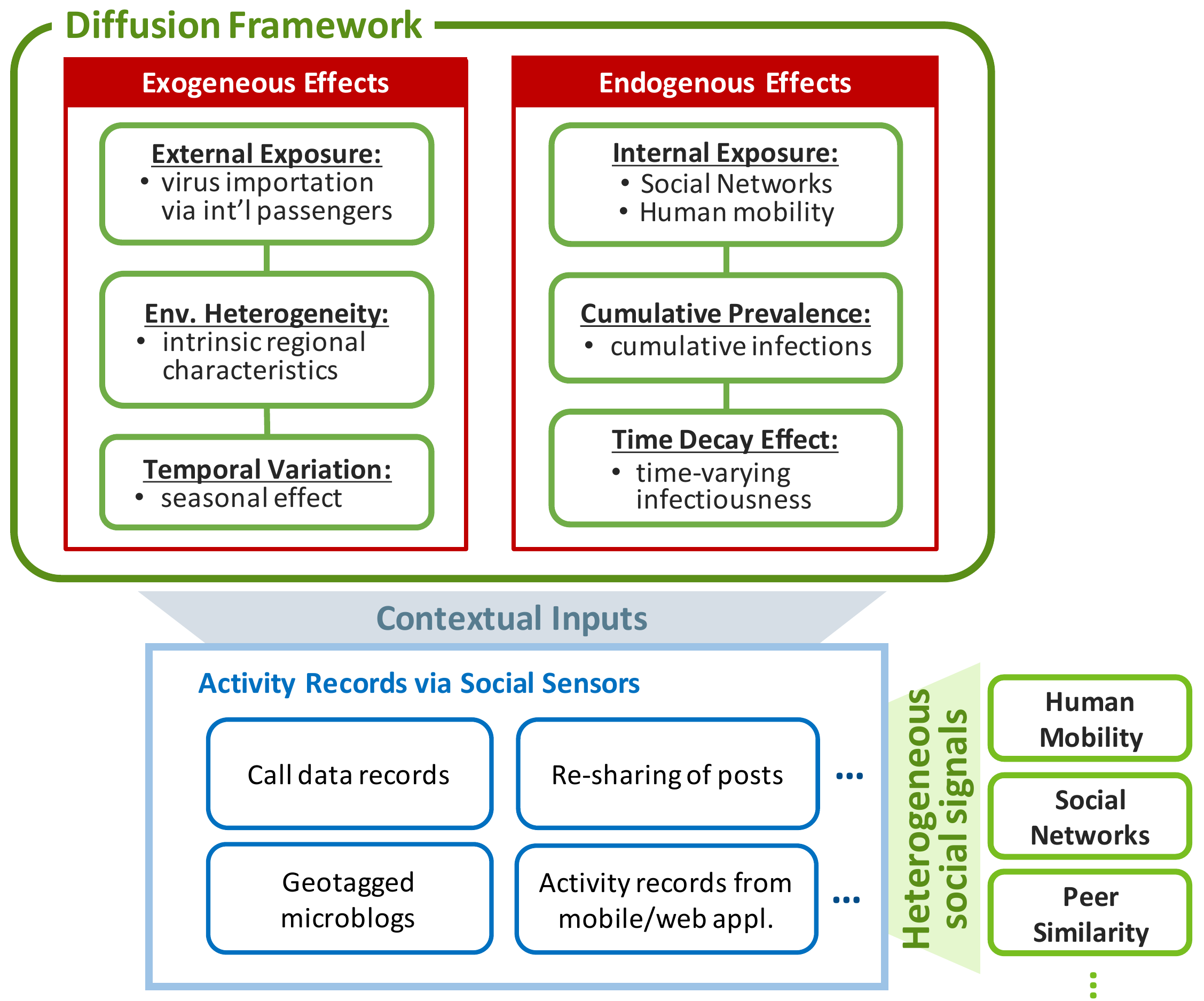}} 
	\end{tabular}	 	
	\caption{Overview of a case study on dengue spread. (a) countrywide outbreaks of an infectious disease over time ($c_i$: $i$-th contagion, $t_i$: the arrival time of $c_i$, $r_j$: $j$-th region where infections occurred, and $N$: total number of infections). (b) countrywide outbreaks in (a) can be decomposed into timelines of regional outbreaks, which can be represented as spatial and temporal point processes. Regions ($r_1, r_2$, and $r_3$) are color-coded, thick colored arrows represent human mobility between regions, and red dashed arrows indicate hidden infection trajectories influenced by human mobility patterns and/or social networks. (c) conceptual diagram of a diffusion framework for the case study. Two major components, exogenous and endogenous effects are presented for brevity, based on the proposed taxonomy in Figure~\ref{fig:taxonomy}. Social sensors~\cite{aggarwal2013social} are lifelogging a broad range of human activities capturing heterogeneous social signals, \ie, different layers of human bursty behavior, which can be inputs to a diffusion framework for providing rich context of underlying diffusion processes.} 
	\label{fig:case_study}
\end{figure*}

The detailed comparisons of the diffusion models based on point processes are presented in Table~\ref{tab:models} from the aspects of the proposed four major universal effects on real-word dynamics in Figure~\ref{fig:taxonomy}. 
Notably, the only applications that consider environmental heterogeneity as part of the exogeneous effects are non-information-based (seismology and criminology), which is likely due to relative environmental independence of information diffusion, particularly online. 
Existing disease-focused studies that model diffusion through point processes also appear to disregard environmental heterogeneity, despite its relevance for larger scale studies. 
Another interesting observation from Table~\ref{tab:models} is that all related studies that consider cumulative prevalence have been related to information diffusion based on nonhomogeneous Poisson processes, which may be linked to the easier traceability of information-based diffusion processes.
As shown in the table, while there has been limited application of diffusion with point processes to animal diseases, the application of diffusion framework to disease spread, particularly for humans, remains significantly under-explored. 
In the next section, we propose a high-level sketch of universal components of the prediction framework for disease diffusion.

%% file: 050casestudy.tex

Predicting disease spread is essential to the prevention and/or control of infectious diseases.
However, it is very challenging, since unlike explicit resharing events between a follower and a followee in social media, the movement of infected people are largely unknown.
Moreover, the recent growth of international travel volumes has brought the outbreaks of exotic viruses, such as dengue, from endemic regions~\cite{gardner2013global,bhatt2013global}, which increases the uncertainty of the infection pathways.
In this context, we focus on the spread of the exogenous virus, dengue, which is a mosquito-borne viral disease transmitted among humans by mosquito vectors~\cite{world2009dengue}, as a case study.

\subsection{Background}
The spread of infectious diseases forms outbreak clusters in both space and time.
As discussed earlier, such space-time clustering of discrete events have been widely observed in diverse research areas, such as aftershock sequences near the epicenter~\cite{ogata1988statistical,ogata1991some,stein1999role} in seismology and burglary sequences along crime hotspots in criminology~\cite{mohler2011self}.
In order to predict the spatial and temporal diffusion of dengue, the four universal components in Figure~\ref{fig:taxonomy} can be readily applied to this case study.

In addition, recent advancements in social sensing technology~\cite{aggarwal2013social} enable the collection of diverse aspects of human social behaviors, \ie, heterogeneous social signals~\cite{kim2017social}, such as online social networking from social media platforms, communications from email or telecommunication systems, human mobility patterns from geo-tagged social media records or mobile phone devices, and so forth.
Such heterogeneous social signals capture a wide spectrum of bursty behaviors, and thus combining different kinds of data sources may capture more realistic and contextual diffusion phenomena.

\subsection{Diffusion Framework with Universal Components}
\label{sec:diffusion_framework}
The universality of real-world diffusion dynamics is important to design a more generalizable framework beyond a research domain.
In this respect, we suggest the following four major factors as universal components of a dengue diffusion framework based on our disclosed universal effects as below.

\begin{myitemize} 
  \item[1)] Exogenous Effects: 
  \begin{myitemize} 
    \item external exposure -- importation of dengue virus via international travelers
    \item environmental heterogeneity -- disease-specific factors such as the distributions of mosquito vectors, temperature, precipitation, and humidity
    \item temporal variation -- time-varying exogenous effects due to seasonal changes in international travel volumes and breeding conditions of mosquito vectors
  \end{myitemize} 
  \item[2)] Endogenous Effects:
  \begin{myitemize} 
    \item internal exposure -- human mobility patterns
	\item time decay -- a cycle of dengue transmission
	\item cumulative prevalence -- cumulative infection counts
  \end{myitemize}
  \item[3)] Diffusion Item:
  \begin{myitemize} 
    \item type -- infectious disease
	\item characteristics -- mosquito vector-borne disease
	\item collection -- nation-wide dengue virus
  \end{myitemize}
  \item[4)] Diffusion Space:
  \begin{itemize}[leftmargin=*]
  	\item[] virus propagation across multiple subregions within a city or country, which can be extended to a global scale
  \end{itemize} 
\end{myitemize}

External virus importation via international travelers from endemic regions may trigger disease spread over domestic social networks, which is accelerated by diverse human mobility patterns across the nation. 
A distinction between exogenous and endogenous effects on the disease spread enables us to quantify each effect and helps to improve the prediction accuracy, which has been relatively neglected in epidemic studies~\cite{colizza2008epidemic,bhatt2013global,gardner2013global}.  

Geographical characteristics conducive to a target disease increase the likelihood of outbreaks, which can be considered environmental heterogeneity (\ie, the intrinsic nature of a region).  
Internal exposure depends on the infection history of connected neighbors in social networks. 
However, collecting nationwide social structures is challenging due in part to privacy issues and dynamically changing social relationships.
Thus, human mobility can be considered for quantifying topological heterogeneity of meta-populations at a macro level~\cite{barrat2008dynamical,colizza2008epidemic,kim2013modeling}, driving population fluxes between regions (\eg, suburbs, cities, states), as one of the heterogeneous social signals.
Regarding the cumulative prevalence, the likelihood of the next diffusion event is proportional to accumulated outbreaks up to the current time, which is counter-balanced by a time relaxation function reflecting the aging effect on dengue transmission.
In terms of diffusion items and space, the dengue virus is a vector-borne infectious disease, whose outbreak risk is rapidly increasing worldwide~\cite{world2009dengue}. 

Figure~\ref{fig:case_study} summarizes the main concept of the case study, which exploits heterogeneous social signals (\eg, human mobility, social networks), captured by lifelogging social sensors, as contextual inputs to our suggested components of a diffusion framework for reflecting more realistic situations and providing rich context of underlying diffusion processes.

Based on our suggested major factors, a cross-regional or cross-metapopulation diffusion model can be a key step toward multinational outbreak risk prediction models.
More importantly, the suggested components are not limited to the dengue case but applicable to other diseases, both vector-borne and infectious.

\eat{ 
\subsection{Research Challenges}
As discussed earlier, heterogeneous social signals enable to capture rich dynamics of diffusion in the real world.
That is, social sensing provides supportive clues to incorporate universal components into a diffusion model, which makes the diffusion framework more generalizable and applicable to varieties of real-world cases than others based only on homogeneous social signals.
Accordingly, there are following research challenges to resolve.
How can we combine different and noisy data sources collected from heterogeneous social signals and obtain significant inputs to a diffusion model? 
How can we avoid a biased estimation caused by a limited sample size? 
Besides the data fusion, a primary question is how to design a general diffusion framework by incorporating the disclosed universal components.
These questions are common to a wide range of diffusion studies, reaching far beyond this dengue case study and demanding interdisciplinary approaches as investigated.
In this respect, we discuss considerations for the design of a general diffusion framework in the next section.
}

%% file: 060discussion.tex

Stochastic models have been key mathematical tools for describing event sequences in time and/or geographical space and accordingly have attempted to incorporate the major factors of clustered inter-event times.

\subsection{Factor Engineering}
Despite the recent advances of diffusion models, there are fundamental questions for the design of a diffusion framework: what does a framework include? How is it organized? What does it explain? Does it have any constraints?
These questions reflect major factors of bursty behavior.
We discuss key design principles of a baseline framework for real-world diffusion dynamics by considering factors as components: (1) factor targeting, (2) factor balancing, (3) factor coverage of diffusion cases, and (4) factor dependency on prior knowledge.

First, {\it factor targeting} is related to a model's capability to provide a rich, but generic interpretation of the underlying diffusion mechanisms.
For instance, microscopic factors increase a model's complexity and make it hard to discover universal and significant effects on diffusion.
In contrast, simple factors rarely provide insights of underlying diffusion processes.
The choice of factors therefore affects a model's performance in terms of complexity and interpretability.
Second, {\it factor balancing} addresses how to separate exogenous from endogenous effects on diffusion and how to balance positive (\eg, preferential attachment) and negative (\eg, time decay) factors within each effect.
For example, the intensity of each node's response to an event is defined as either a multiplicative process of exogenous and endogenous topological effects~\cite{gomez2013modeling}, a simple additive process of innovation and imitation within a homogeneous population~\cite{bass1969,bass2004comments}, a multiplicative process of detailed endogenous factors (positive and negative) without exogenous effects~\cite{wang2013quantifying,shen2014aaai,zhao2015seismic}, or a combination of those additive and multiplicative processes~\cite{myers2012information,luu2012modeling,kim2013entropy,iwata2013discovering,farajtabar2015coevolve}.
Third, {\it factor coverage} implies the applicability of a model to real-world cases.
A single model cannot explain all diffusion cases in the real world, but it needs to be designed to cover real diffusion cases of a target domain as much as possible.
Finally, {\it factor dependency} is related to factor targeting and coverage.
Strong dependency on specific domain knowledge will fail to define a general model, which limits the coverage and thus interpretability of a diffusion model.

\subsection{Research Challenges}
As discussed in the previous section, heterogeneous social signals enable the capture of rich diffusion dynamics in the real world.
That is, social sensing provides supportive clues to incorporate universal components into a diffusion model, which makes the diffusion framework more generalizable and applicable to a broad range of real-world scenario, in comparison with models based only on homogeneous social signals.

Accordingly, these are the following research challenges to resolve. 
How can we combine different and noisy data sources collected from heterogeneous social signals and obtain significant inputs to a diffusion model? 
How can we avoid a biased estimation caused by a limited sample size? 
Besides the data fusion, a primary question is how to design a general diffusion framework by incorporating the disclosed universal components.
These questions are common to a wide range of diffusion studies, demanding interdisciplinary approaches to be investigated.

%% file: 070conclusion.tex

Collective bursty behaviors have been observed across different real-world domains, exhibiting clustered interevent times.
That is, emergent bursts provide supportive clues of underlying diffusion processes, which can be modeled with point process approaches for defining random events in time and/or space and for better understanding the diffusion dynamics of individual items in a real world setting.

In this context, we reveal common major factors of diffusion processes and propose a taxonomy of universal effects on real-world diffusion dynamics across disciplines (\eg, computer science, social science, statistical physics, seismology, criminology, marketing, finance, and epidemics).
As a case study, we suggested the four major factors of dengue spread as components of a disease diffusion framework.

We expect that our work presented here will shed light on modeling a more general diffusion framework by incorporating the disclosed universal effects as contextual components into the framework.
An interesting direction for future work would be to model disease spread by incorporating the suggested factors in our case study.